\documentstyle[12pt,fleqn,espcrc1,epsfig]{article}
\title{Quark Matter '99 --- Theoretical Summary: What Next?}

\author{Berndt M\"uller\\
        Department of Physics, Duke University, Durham, NC 27708-0305}

\begin{document}

\maketitle

\section*{Introduction}

Having been asked by the organizers of this conference to summarize
what we learned here from a theoretical perspective -- to be followed
by Reinhard Stock's experimental summary -- I will first review the
three broad areas where major progress has been reported: The phase
structure of strongly interacting matter, the properties of matter
at the instant when it freezes out into individual hadrons in the 
final stage of the expansion of the hot fireball, and the status of
the main signatures of the formation of a quark-gluon plasma. In the
final section I will offer some thoughts about what should be done
next, both in the experiemntal and the theoretical arena.

\section*{The Phases of QCD}

Insight into the phase structure that quantum chromodynamics imposes
on the world continues to be full of surprises. This issue, which lies
at the basis of the experimental relativistic heavy ion programs around 
the world was the subject of several talks at this meeting.

Di Giacomo reminded us that quarks are, indeed, confined. Arguments
based on cosmological evidence show that free quarks are at least
$10^{-15}$ less abundant than expected. This implies that the
Gibbs factor for a free quark at the confinement phase transition
must have been limited by $\exp(-E_{\rm q}/T_{\rm c}) \ll 10^{-15}$
or $E_{\rm q} > 10$ GeV. This number is so large (on the QCD scale
$\Lambda$) that one expects a general principle at work: a kind of 
dual Meissner effect. If the QCD vacuum is a color-magnetic
superconductor, i.e. if color-magnetic monopoles form a condensate
in our normal world, the Meissner effect would naturally explain
the absence of any color-nonsinglet excitation around us \cite{Mand76}.

Color-magnetic monopoles do not exist as fundamental objects in the 
QCD Lagrangian, but the Yang-Mills equations allow for extended,
though unstable, topological excitations with the quantum numbers
of magnetic monopoles in a [U(1)]$^{N-1}$ subspace of the SU($N$)
gauge group. These excitations can be identified in the lattice
gauge theory by abelian projection of the link variables after
choosing an appropriate gauge. Thus, any attempt to prove the
existence of a monopole condensate is inherently gauge dependent.
However, by repeating the study in several different gauges, the
general nature of the result can be checked. 

Di Giacomo showed us the results of lattice simulations
investigating the dependence of the quantity 
$\rho = d(\ln\langle\mu\rangle)/d\beta$,
$\langle\mu\rangle$ being the monopole condensate, on
the inverse temperature $\beta$. He presented impressive evidence
\cite{DelDeb95} that the quantity behaves as 
$\rho \propto (\beta_{\rm c}-\beta)^{-\nu}$ with an exponent 
$\nu$ indicating a second-order phase transition in SU(2)
and a first-order transition in SU(3). These results constitute
strong evidence that color-magnetic monopole condensation
is the cause of color confinement in non-abelian gauge theories 
without dynamical quarks. It will be very interesting to see how
the result extends to real QCD.

A quite different picture of the QCD phase transition was presented
to us by Helmut Satz, who argued that QCD undergoes a percolation
phase transition \cite{Satz98} at $T_{\rm c}$. In the
poster session, Fortunato showed that the percolation idea can be
made precise by studying the domain structure of the sign of the
Polyakov loop, i.e. the trace of the timelike Wilson line
\begin{equation}
\sigma(x) = {\rm sgn} \left( {\rm tr} (e^{-\int A d\tau}) \right) ,
\end{equation}
in the SU(2) gauge theory on an euclidean lattice. The numerical
results demonstrate cluster percolation exactly at the critical
temperature identified in the conventional way. Again, this study
needs to be extended to SU(3) and to the theory with dynamical
quarks before one can safely claim relevance to the real world.

However, it is tempting to speculate that cluster percolation is
related to the Hagedorn transition in string theory, which is known
to have the nature of a Kosterlitz-Thouless transition with
topological defects proliferating on the world-sheet of the string
wrapped around the periodic euclidean time dimension \cite{Kogan}.
In terms of simple concepts, at the critical temperature the dilute 
string gas assembles into a dense web of interconnected strings that 
permeates the whole of space. It would be interesting to see whether 
this analogy could be made more precise in some of the exactly solvable
models of supersymmetric gauge theories derived from superstring
theory.

Rajagopal, in his talk, gave an overview of the recent developments
that have made the QCD phase diagram even more interesting. We now
understand, as first pointed out seriously by Bailin and Love 
\cite{BL84} in the mid-1980s, that di-quarks condensate at large 
baryon density when the temperature is not too high \cite{ARW98}.
The high-density ground state of 
QCD is thus a color superconductor while, as discussed before, the
QCD vacuum is most likely a dual color superconductor. Depending on
the masses of the light quarks, the detailed properties of the 
high-density phase vary, but in some cases may be quite exotic:
rotational symmetry as well as parity may be spontaneously broken!
Most of these studies are based on phenomenological models of the
quark-quark interaction, but the existence of Cooper pairing between
quarks can be proven rigorously in the very-high density limit
where one-gluon exchange becomes perturbatively calculable.
Unfortunately for our field, because of their restriction to ``low''
temperature but high density, these phenomena probably cannot occur
in heavy ion collisions; but they remain of astrophysical interest
and may occur in the dense, cold cores of collapsed stars.

We were also reminded that symmetry arguments suggest the presence
of a critical point somewhere along the QCD phase boundary, which is
the remnant of the tri-critical point separating the regions of first-
and second-order phase transition in a world with two massless quark
flavors \cite{Steph}.
Because lattice calculations at finite baryon density are 
still impossible, we do not know exactly where this critical point is, 
but we can look for it experimentally by searching for anomalously
large fluctuations caused by the critical fluctuations in its
vicinity. Since no evidence for such fluctuations has been observed
at the SPS, where the quark chemical potential $\mu\approx 90$ MeV is
already quite low, it is reasonable to expect that the critical point
could be located somewhere between the conditions accessible at the
SPS and those created in nuclear collisions at the AGS ($\mu\approx
180$ MeV). If critical fluctuations persist within a sizable region
around the critical point (models suggest for $|\mu-\mu_{\rm c}| <
0.2 \mu_{\rm c}$), experiments at just a few beam energies may be
sufficient to locate the critical point.

Our understanding of the fully developed quark-gluon plasma also has
recently made significant progress. Several groups of theorists have
shown that the picture of the QGP as a plasma of weakly interacting 
quasi-particles with effective masses $m_{\rm g}^{\rm eff},
m_{\rm q}^{\rm eff} \sim gT, g\mu$ accounts remarkably well for
the thermodynamic properties of the QGP phase \cite{Peshier}.
Comparisons with the
results of lattice gauge theory for $\mu=0$ indicate that this simple
description may work until very close to $T_{\rm c}$. Since this
analytic technique can be easily extended to nonzero $\mu$, it could
provide a useful description of the full QCD phase diagram for
phenomenological purposes.

There has been much progress in QCD transport theory recently,
which was not reported at this conference. The extension of the
kinetic mean-field theory of non-abelian plasmas to a Boltzmann
equation including collisional effects, as well as the formulation 
and numerical solution of the mean-field theory on real-time 
lattices are especially noteworthy \cite{Manuel}.
Other interesting work concerns the description of dynamical color 
screening in nuclear collisions within the framework of Geiger's
parton cascade model \cite{Dinesh}. Independently, the perturbative 
QCD calculations shown by Eskola in the RHIC predictions session, 
predict perturbative saturation of partons in Pb+Pb at RHIC for a 
cut-off momentum $p_0\approx 1$ GeV/c. In such a picture, the entire 
transverse energy generated in the collision could be produced by 
minijets! While this may be exaggerated, it indicates that the idea 
that a quark-gluon plasma is created by partonic processes is not 
totally unreasonable.

Obviously, the challenge is to put such a picture on firm ground by
developing a formalism of parton transport with medium effects built
in self-consistently. Work in this direction is proceeding steadily
if slowly \cite{Makhlin}. In the meantime, the theoretical concept of 
gluon radiation by rapidly moving color charges, developed by the
Minnesota group and others, has been successfully applied to 
electrodynamic processes. The fast-moving, peripherally
colliding nuclei are viewed as sheets of electric charge along the
intersecting light-cones which generate a flash of electromagnetic
field energy when they briefly meet. This pictures allows for 
calculations of important processes, such as electron-positron pair 
creation with and without capture of the electron, ionization, and 
Coulomb fission of the colliding nuclei, to all orders in the nuclear
charge in the high-energy limit \cite{Baltz}.

\section*{The Hadronic Freeze-Out}

Tremendous experimental and theoretical progress has been made 
recently towards the full characterization of the conditions at
the moment of the final freeze-out of hadrons emitted from the
nuclear reactions, both in the AGS and SPS energy domain. We have
almost complete data on the spectra and yields of a wide variety
of hadrons, including $\pi^{\pm,0}, K^{\pm,0}, \phi, p, \Lambda,
\Xi^-, \Omega^-$, and their antiparticles. Pair correlations
of charged and neutral pions, positive kaons, protons, and even
$\Lambda$-hyperons have been measured. We have seen remarkable
data, especially from the AGS, of yields of light (anti-)nuclei:
from d up to $^7$Li, $\overline{\rm d}$, $\overline{^3\rm He}$,
and the hypernucleus $^3_{\Lambda}$H.

As was pointed out by Wiedemann in his review lecture, a highly
consistent picture is emerging from these measurements \cite{Urs}.
The single-particle momentum spectra, the $p_T$-dependence of the
pair correlations, and the fragment yields all can be explained 
by a freeze-out from a thermal, dilute hadronic fireball with
a final temperature (at the SPS) around $T_{\rm f}\approx$ 100 MeV,
an average transverse flow velocity $\langle\beta_{\rm f}\rangle
\approx 0.55$, and a baryon chemical potential $\mu_{\rm f}\approx
400$ MeV. The transverse rms radius of the fireball at freeze-out 
is approximately 10 fm; and one obtains a lower limit for the 
average freeze-out time of about 6--8 fm/c after impact, spread 
over a rather short period of 2--3 fm/c.\footnote{The mean 
freeze-out time cannot be unambiguously deduced from the data and 
is model dependent.} Thus, the freeze-out occurs quite suddenly, 
after the fireball has expanded considerably from its original size.

Systematic differences between various fits using different
parametrizations of the freeze-out geometry still need to be
resolved, but it is clear that the required data are now
available. These high-quality data impose severe constraints
on any phenomenological model. For example, Bose-Einstein
correlations of identical pions show \cite{Ferenc} that the 
average density of phase-space occupation of pions at freeze-out 
cannot much exceed the value $\langle f\rangle = 0.3$.

The data also prove that the chemical composition of the expanding
hadronic matter is frozen much earlier. Equilibrium fireball model
fits give values \cite{PBM}
$$
T_{\rm ch} = 170 \pm 10 {\rm MeV},\qquad 
\mu_{\rm B} = 270 \pm 25 {\rm MeV}
$$ 
for Pb+Pb at the SPS and 
$$
T_{\rm ch} = 125 \pm 10 {\rm MeV},\qquad 
\mu_{\rm B} = 540 \pm 25 {\rm MeV}
$$ 
for Au+Au at the AGS. The hadronic yield data for the AGS are
not quite as complete, but the obtained values are compatible
with the nuclear cluster data. 

It remains contentious whether there is evidence for chemical 
non-equilibrium at mid-rapidity.
Although it is clear that the chemical composition changes 
as function of the rapidity $y$, especially at the AGS, this
effect may be at least partially described by $y$-dependent
thermodynamic parameters. To avoid such complications, many
chemical analyses only consider phase-space integrated hadron
yields. This procedure tends to obscure non-equilibrium effects.
Some theorists still argue vehemently that such effects exist and 
constitute important evidence \cite{Rafelski}. I believe that the data 
now are sufficiently precise to settle this issue once and for
all, and to establish the separation of chemical and thermal 
freeze-out objectively and unambiguously.
Microscopic models of the late stage of the expansion of the
hot hadronic fireball could serve to help identify distinctive
signatures of non-equilibrium \cite{Bass98a}. One specific case 
is the ``early'' freeze-out of the $\Omega$ hyperon seen by 
experiment WA97 and its prediction by hadronic cascade models 
\cite{Sorge}.

This brings me to the topic of microscopic theoretical models
for nuclear reactions in the SPS energy domain. Hadronic 
cascade models, some with mean-field interactions, have succeeded 
in reproducing the gross and many detailed features of the 
nuclear reactions. They have become indispensible tools for
experimentalists who wish to identify interesting features in
their data or make predictions to plan new experiments.
However, the general success of these models can easily lead
to misconceptions. As detailed comparions with the new high
precision data reveal, {\em none} of the existing hadronic 
cascades really works correctly at SPS energies. As Antinori
impressively demonstrated in his talk, the comprehensive data 
of WA97 on (anti-) hyperon yields rule out even those models 
that have been specifically ``tuned'' by the addition of novel 
mechanisms \cite{WA97,Antinori}.  Moreover, all model descriptions 
based on hadronic dynamics are fundamentally inconsistent at high 
densities, calling their application to describe collisions among 
heavy nuclei at the SPS into question.

This is revealed quite dramatically when one asks, which fraction 
of the energy is contained in ``standard'' hadrons, i.e. those 
with entries in the Particle Data Booklet, and which fraction is 
temporarily stored in more fictitious components, such as 
pre-hadronized strings. Bass and coworkers have studied this 
question within the framework of the UrQMD model \cite{Bass98}. 
They find that up to a time of 8 fm/c most of the energy density resides 
in strings and other high-mass continuum states that have not fully decayed. 
Clearly, we know very little from first principles about the physical
properties of these objects even when they occur in isolation, 
and practically nothing about their interactions (or even their 
existence) in a dense environment. Any skeptical scientist must
conclude that the application of these models to the early 
phase of a Pb+Pb collision at the SPS is highly speculative.

While this insight is not altogether new, it raises the question
whether any microscopic hadronic model can sensibly be used to 
describe nuclear collisions at the SPS in their entirety. Those 
who believe so need to present arguments that clarify the meaning 
of hadronic cascades under such conditions and answer to
the question how the theoretical consistency of their models 
can be assessed. Until agreement on this issue has been reached,
I suggest that theorists should refrain from:
\begin{itemize}
\item trying to ``fine-tune'' existing codes to describe all
      details of the final state and claiming that the success 
      has any physical implication;
\item writing another, ostensibly ``better'' code;
\item applying purely hadronic cascade models to nuclear 
      collisions at RHIC energies.
\end{itemize}
Instead, experimentalists should insist that codes based on such
models must contain a flag that automatically generates a warning 
message when the limit of credible applicability of the hadronic 
cascade is reached (e.g. when strings begin to overlap), or when 
the final result depends essentially on fictitious components 
of the model that are not based on experimental evidence 
(color ropes or quark droplets, interacting baryon junctions,
etc.). The users of such codes would then know that the
result is not a prediction of known physics but, at least
in part, the based on speculative ideas of its author.

In the meantime, I would like to suggest that we begin to
use the existing hadronic cascade codes more seriously as 
marvelous tools for the description of the freeze-out process. 
These models comprise a large body of information about
reactions among hadrons in a dilute environment where the
physics is dominated by two-body interactions involving
on-shell hadrons and known hadronic resonances. A big step
in this direction was recently taken by Bass in collaboration
with several other young theorists, when he used the UrQMD
code to model the dynamics of the late hadronic phase that
follows the evolution of dense matter described by a parton
cascade or a hydrodynamic model \cite{Bass99}.

The existing hadronic cascade codes are ideally suited for
systematic studies of the chemical and thermal freeze-out
of hadrons when the dense phase of the nuclear reaction is
over. They are vastly superior to the simple models of
hadrochemical equilibration that were used earlier to explore 
the possible hadronic mechanisms that could lead to strangeness 
enhancement and equilibration in a hadronic gas \cite{Koch}.
These studies neglected higher mesonic and baryonic resonances
and usually only considered small deviations from thermal
equilibrium. Significant improvement over these early
calculations is possible and would be highly desirable
for a complete analysis of the implications of the wealth
of freeze-out data now available.

\section*{QGP Signatures}

Of the many probes for the high-density phase of QCD in
nuclear collisions, only four robust signatures have
withstood the experimental tests:
{\em A:} Flow (directed, radial, elliptical);
{\em B:} Flavor equilibration;
{\em C:} $J/\psi$ suppression;
{\em D:} $\rho$-meson broadening or disappearance.
The data from the SPS heavy ion experiments presented at this 
meeting have brought impressive and sometimes spectacular evidence 
for signatures {\em A--C}. Signal {\em D} has also been firmly
established, and can be expected to reach a similar state 
of maturity when data taken with the upgraded CERES
detector will become available. I will now discuss the
four signatures, in turn.

{\em A. Flow:} One of the most exciting new messages of
this meeting was the observation of the softening of the
equation of state already in the AGS energy range, around
3--4 GeV/u, in Au+Au collisions \cite{E895}. Microscopic 
models predict that the highest baryon density reached at 
this energy is about 4--6 times normal nuclear density. 
The precise transition from out-of-plane to in-plane 
elliptical flow could be explained if baryon dense matter 
makes the transition from a stiff equation of state to a 
much softer one in the vicinity of this point \cite{Daniel}. 
This phenomenon is seen in a wide range of
microscopic models. It should be carefully studied whether
the softening can be accounted for by the onset of copious
baryon resonance production, or whether it requires an
additional change in the structure of baryonic matter.

Many descriptions of baryon-dense matter predict
the onset of chiral symmetry restoration for precisely this
density range (at $T\approx 100$ MeV). The question is, how
can this issue be further studied? One could try to obtain
more systematics, e.g. the dependence of the phenomenon on
the size of the colliding nuclei. Also, it may be time to
reconsider the possibility of a lepton-pair experiment
in the AGS energy domain to search for the disappearance 
of the $\rho$-resonance in the lepton-pair spectrum.

\begin{figure}[htb]
\vfill
\centerline{
\begin{minipage}[t]{.65\linewidth}\centering
\mbox{\epsfig{file=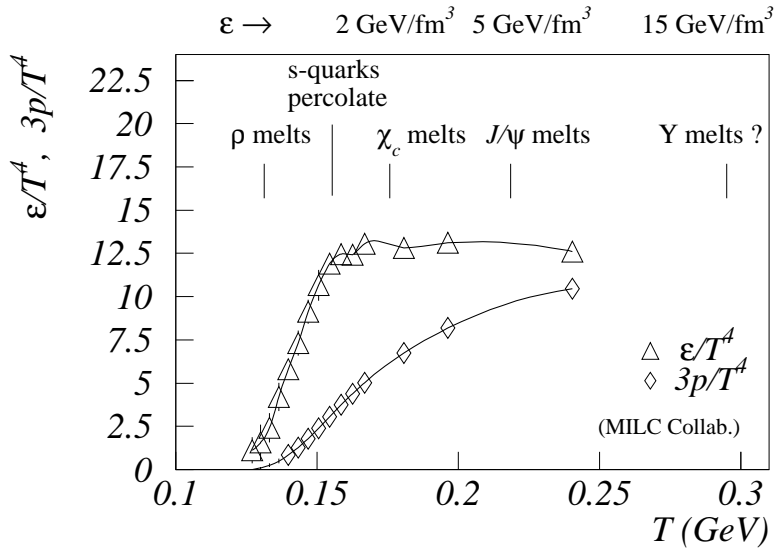,width=.99\linewidth}}
\end{minipage}
\hspace{.04\linewidth}
\begin{minipage}[b]{.31\linewidth}
\caption{Lattice-QCD equation of state and QGP signatures.}
\label{fig1}
\end{minipage}}
\vspace{-2mm}
\end{figure}

{\em B. Strangeness:} The data obtained at the AGS and SPS
leave no doubt that strangeness is enhanced throughout
the baryonic sector. Nagle \cite{Nagle} reported the value 
3.5 for the ratio $\bar\Lambda/\bar p$ for Au+Au at the AGS
(expt. E864), and Antinori \cite{Antinori} showed that the new 
p+Be run at the SPS has confirmed the enhancement by a factor 15 
of the $\Omega$ hyperon in Pb+Pb with respect to p+$A$ collisions
(expt. WA97). Few thought such enhancements possible when
they were first predicted in the mid-1980s in the framework
of models based on quark-gluon plasma evaporation. In order
to realize the full power of these experimental results, new
studies of flavor equilibration in the hadronic phase using
modern cascade models are urgently needed.

{\em C. Charmonium:} The new data presented by NA50 clearly
show that the $J/\psi$ state is nontrivially suppressed in
Pb+Pb collisions, with the suppression continuing to increase
up to the highest values of $E_T$ accessible in Pb+Pb 
\cite{NA50}. Problems with multiple interactions at high $E_T$
that contaminated earlier data have been resolved. The use
of minimum bias data to normalize the $J/\psi$ measurements
now permits NA50 to fully exploit their high statistics.
Unfortunately, the $J/\psi$ issue is  
plagued on the theoretical side by an abundance of ``cheap''
models that prove nothing. As Redlich showed in his talk,
the state of the theory of interactions between $J/\psi$
and light hadrons is embarrassing \cite{Redlich}.  Only
three serious calculations exist (after more than 10 years
of intense discussions about this issue!), and their results
differ by at least two orders of magnitude in the relevant
energy range \cite{JPSI}. There is a lot to do for those who 
would like to make a serious contribution to an important 
topic.

{\em D. Low energy lepton pairs:} What is possible when the
power of modern hadron theory is brought to bear on a
related subject, was clearly demonstrated at this meeting
in Rapp's and Eletsky's talks about in-medium interactions
of the $\rho$-meson and applications to low-mass lepton
pairs from heavy ion interactions at the SPS. Various
different approaches have come to a remarkable convergence
\cite{RHO}.
It has become clear that the $\rho$-resonance is strongly
broadened in hot, baryon-rich matter, while its centroid
barely moves. Under SPS freeze-out conditions, predicted
values \cite{EIK98} are $\Delta\Gamma_{\rho} \approx 300$ MeV,
$\Delta m_{\rho} \approx 10$ MeV. It has also been shown
that nucleons are more efficient in this respect than
pions.

Rapp also pointed out that the hadronic correlator in the
$\rho$-meson channel ($1^{--}$)
$$
C_{\mu\nu}(p)=\int dx e^{ipx}\langle j_\mu(x)j_\nu(0)\rangle
$$
rapidly approaches the form expected from perturbative QCD
when $T$ and $\mu_B$ approach the critical line separating
hadron from quark matter. $\rho$ broadening and $\rho-a_1$  
mixing combine to generate a spectral function in the photon
channel resembling that of free $q\bar q$ annihilation
\cite{Rapp}.
The similarity might be even closer if interactions among
the quarks were included, especially the effective mass of
the quasiparticle mode with quark quantum numbers in the
dense, hot medium.

Let me finally point out that all observations are consistent
with a succession of several stages of dissolution of 
hadronic states as the energy density increases from light
to heavy nuclear collision systems at the SPS (see Figure 
\ref{fig1}). As the critical conditions 
($T_{\rm c},\mu_{\rm c}$) are approached from below,
first the $\rho$-meson dissolves and strange quarks begin to
percolate freely among hadrons. The full population of strange
quark phase space within a few fm/c probably requires the
activation of gluonic excitations. These also lead to the
disappearance of the weakly bound $\chi_c$ state which accounts
for the anomalous $J/\psi$-suppression in all but the most
central Pb+Pb collisions. Finally, the gluon density becomes
sufficiently high and energetic to dissolve the $J/\psi$
itself in the innermost core of the Pb+Pb fireball. The
energy density required is predicted to be around 5 GeV/fm$^3$.
This value may just be reached in the most central Pb+Pb 
events. It is time for the theory community to focus 
its efforts on confirming (or refuting) this scenario,
and to suggest new experiments or ways of data analysis
that can help us to reach a conclusion.

\section*{What next?} 

The organizers of this conference asked me to say a few
words about what, in my opinion, should be done next to
resolve the remaining issues that stand in the way of
a clear-cut interpretation of the data. Here are some
thoughts:
\begin{itemize}

\item A high-resolution study of the low-mass lepton pair
spectrum in Pb+Pb is needed that identifies the $\omega$
resonance and confirms that the enhancement below the $\rho$
is accompanied by a suppression of the $\rho$ itself. This
investigation is under way with the upgraded CERES detector.

\item The $J/\psi$ and flavor signals need measurements
in smaller symmetric nuclear systems (Ag+Ag, Ca+Ca ?).
This would allow to resolve discrepancies between theory
and the NA50 data points at low $E_T$ where systematic
errors become serious, and should help locate the onset
of strangeness enhancement and flavor equilibration.

\item The flow and strangeness signals, as well as the
quest for critical or other nontrivial fluctuations need
data from collisions at lower beam energies. Again, this
is under way with the run at 40 GeV/u later this year.

\item The indications of an enhanced open-charm background
in the di-muon spectrum reported by NA50 calls for a precise
measurement of open charm production. Although the 
enhancement is not anticipated by theory, it needs to be
either confirmed or firmly refuted experimentally, to
rule out false conclusions about the suppression of the
$J/\psi$. Fortunately, the NA50 collaboration has proposed
an experiment that could measure the D-meson yield, and a
similar proposal is in preparation by the NA49 collaboration.

\item The flow data from the lower AGS energy range (2--6
GeV/u) call for measurements with smaller nuclei, which
would allow to change the baryon density at fixed kinematics.
The data also reaffirm the need for a lepton-pair experiment 
in this energy range.

\item The possible termination of the SPS heavy ion program
in two years is of serious concern, because it would virtually
preclude the careful measurement of the excitation function
of the hadronic observables below the maximal SPS energy. The
new data from the AGS, covering the energy range between the
Bevalac/SIS and the highest AGS energy, many of which
were first presented at this conference, have impressively
demonstrated the importance of such a program. 

If it turns out to be impossible to perform a systematic 
exploration of the energy range between the top AGS energy and 
the top SPS energy at CERN, one should seriously consider 
whether such a fixed target program could not be initiated 
at RHIC. As Rafelski and Uggerh{\o}j  have begun to point out, 
a small fraction of the RHIC beam could be 
extracted by means of crystal channeling \cite{RU99}.
Since there are already plans at RHIC to use channeling
to ``clean up'' the intersecting beams at one intersection,
this might be an inexpensive way to provide a low-intensity
beam for fixed-target experiments. Obviously, because of the
low event rates, such a program would have to focus on 
hadronic observables.

\item Finally, let me make a few remarks that concern the
theory community interested in relativistic heavy ion physics.
It is important that we seriously address the challenge 
posed by the new high-quality data, from the AGS as well as
the SPS. The experiences made in other areas of nuclear and
high energy physics, telling us how theory can be used in
conjunction with experimental data to create lasting scientific
progress, may provide useful aid. Here are some suggestions.
Theorists should:
\begin{itemize}
\item calculate {\em carefully} what they can;
\item be mindful of the limits of validity of their 
      favored approach;
\item take recourse to general principles (symmetries,
      effective theories, etc.) where microscopic
      approaches are not feasible.
\end{itemize}
One last thought:
If we want to establish the existence of a new phase (or
phases) of QCD where quarks and gluons play a more direct
role as effective degrees of freedom than in our hadronic
world, we cannot hope to do so without direct reference
to the beautiful theory of quantum chromodynamics. Much
progress is being made in applying QCD to processes that
are of interest to relativistic heavy ion physics, and it
would be a mistake to ignore QCD as we make the
transition to the RHIC era.
\end{itemize}

\section*{Acknowledgements}

This work was supported in part by the U.S. Department of
Energy under grant DE-FG02-96ER40495. I thank S.A. Bass 
for comments on a draft of the manuscript.

\end{document}